\documentclass[useAMS,usenatbib]{mn2e}
\usepackage{epsf}

\title[Properties of host haloes of Lyman-break galaxies 
and Lyman-alpha Emitters]
{Properties of host haloes of Lyman-break galaxies 
and Lyman-alpha Emitters from their number densities and 
angular clustering}

\author[T. Hamana et al.]
{Takashi Hamana$^{1,2}$, Masami Ouchi$^3$, Kazuhiro Shimasaku$^{3,4}$,
Issha Kayo$^5$ \and and Yasushi Suto$^{4,5}$\\
$^1$ Institut d'Astrophysique de Paris, 98bis Boulevard Arago, 
F 75014 Paris, France\\ 
$^2$ National Astronomical Observatory of Japan, Mitaka, 
Tokyo 181-8588, Japan\\
$^3$ Department of Astronomy, School of Science, The University of Tokyo, 
Tokyo 113-0033, Japan\\
$^4$ Research Center for the Early Universe (RESCEU), 
School of Science, The University of Tokyo, Tokyo 113-0033, Japan\\
$^5$ Department of Physics, The University of Tokyo, 
Tokyo 113-0033, Japan}

\date{Accepted ******; Received ******; in original form 2003 July 10}
\pagerange{\pageref{firstpage}--\pageref{lastpage}} 
\pubyear{2003} 
\volume{000}

\begin{document}

\label{firstpage}

\maketitle

\begin{abstract}
We explore empirical relations between three different populations of
high-redshift galaxies and their hosting dark halos employing the halo
model approach. Specifically we consider LBGs (Lyman-break galaxies) at
$z\sim4$ and at $z\sim5$, and LAEs (Lyman-Alpha emitters) at $z\simeq
4.86$, all from the Subaru Deep Field survey extending over an area of
about 600 arcmin$^2$.  We adopt a halo occupation function (HOF)
prescription to parameterize the properties of their hosting halos and
the efficiency of halo-dependent star formation.  We find that the two
LBG samples are well described by the halo model with an appropriate
HOF.  Comparing the model predictions with the observed number densities
and the angular correlation functions for those galaxies, we obtain
constraints on properties of their hosting halos.  
A typical mass of hosting halos for LBGs 
is $5\times 10^{11}h^{-1}M_\odot$ and the expected number of
LBGs per halo is $\sim 0.5$, therefore there is an
approximate one-to-one correspondence between halos and LBGs.
We also find a sign of the minimum mass of LBG hosting halos 
decreasing with time, although its statistical significance is not strong. 
We discuss implications of these findings on the star formation history 
of LBGs.
On the other hand, for LAEs, our simple HOF
prescription fails to reproduce simultaneously the observed angular
correlation function and the number density.  In particular, a very high
amplitude of the correlation function on scales larger than 120 arcsec
cannot be easily reconciled by the HOF model; a set of parameters which
account for this high correlation amplitude on large scales predict
either excessive clustering on small scales or a much smaller number
density than observed.  While this difficulty might imply either that
the distribution of LAEs within hosting halos differs from that of dark
matter, or that the strong large-scale correlation is due to the
existence of an unusual, large overdense region, and so the survey
region is not a representative of the $z\sim 5$ universe, the definite
answer should wait for a much wider survey of LAEs at high redshifts.
\end{abstract}


\section{Introduction}

Multi-band color selection techniques (Steidel et al.~1996; 1998; Madau
et al.~1996; Cowie \& Hu 1998; Ouchi et al.~2001, 2003a) have
significantly increased high-redshift galaxy catalogs both in quality
and in size.  Since those high-$z$ galaxies are naturally expected to be
progenitors of the present-day galaxies, their statistical analysis is
of fundamental importance in understanding the formation and evolution
history of galaxies.  Actually, recent large high-$z$ galaxy catalogs
allow one to estimate their luminosity functions and spatial correlation
functions at different $z$ with a reasonable accuracy (e.g., Steidel et
al.~1999; Adelberger et al.~1998; Giavalisco \& Dickinson 2001; Ouchi et
al.~2001, 2003a; Porciani \& Giavalisco 2002).

In the standard scenario of structure formation, it is thought that dark
matter halos are first formed via the gravitational amplification of
initial small density fluctuations.  Subsequently baryonic gas trapped
in a gravitational potential of the dark matter halo becomes
sufficiently dense to cool and to form stars, and such initially small
systems experience repeated mergers to form larger galaxies.  The
formation process of halos is determined by gravity only, thus it is
well understood from $N$-body simulations and also from simple but
relevant analytical approximations such as the Press-Schechter model
(Press \& Schechter, 1974) and its extensions (Bond et al.~1991; Bower
1991; Lacy \& Cole 1993; Mo \& White 1996; Sheth, Mo \& Tormen 2001;
Sheth \& Tormen 2002).  The formation of {\it galaxies}, on the other
hand, involves many complicated processes including hydrodynamics,
radiative processes, star formation, and supernova feedback, and thus it
is hard to solve in a reliable manner even using state-of-the-art
numerical simulations.  Therefore, a simplified or empirical model that
describes essential physical processes is still a valuable tool to
understand basic elements of the formation and evolution of galaxies.

This is why we attempt in what follows to apply an empirical
parameterized model that relates the galaxy number distribution to mass
of the hosting dark matter halo. To be more specific, we explore
statistical relations between two populations of high-$z$ galaxies and
their hosting dark matter halos: (i) Lyman break galaxies (hereafter
LBGs) which are isolated in a color--color diagram due to their UV
continuum depression (Steidel et al.~1996; 1998; Madau et al.~1996;
Adelberger et al.~1998; Ouchi et al.~2001) and (ii) Lyman $\alpha$
emitters (LAEs) which are identified due to their strong Lyman $\alpha$
emission from narrow band imaging (Cowie \& Hu 1998; Hu, Cowie \&
McMahon 1998).  We consider three catalogs generated from the Subaru
Deep Field survey data (Ouchi et al.~2001, 2003a; Shimasaku et
al.~2003); LBGs at $z\sim4$, LBGs at $z\sim5$ and LAEs at $z\simeq
4.68$.

The major purpose of our current analysis is twofold; the first is to
clarify the difference of hosting halos for LBGs and LAEs.  It is well
known that these two populations exhibit different statistical
properties, including the fact that LAEs are in general fainter and
smaller, and are more strongly clustered than LBGs (Ouchi et al.~2003a).
These differences should retain information of their formation processes
as well as environmental effects.  The second is to examine the
difference between LBGs located at different redshifts ($z\sim 4$ and
$\sim 5$). Combined with the previous analysis of LBGs at $z\sim 3$ by
Moustakas \& Somerville (2001) and Bullock et al.~(2002), our results
would provide better understanding of the evolution of LBGs.

For those purposes, we employ the halo approach that attempts to model
the spatial distribution of galaxies in a parameterized fashion.  The
key quantity in this approach is the halo occupation function (HOF) that
describes statistical relations between galaxies and their hosting
halos.  We adopt a simple form for HOF motivated by the semi-analytic
galaxy formation models (Benson et al.~2000; Kauffmann et al.~1999) as
well as by hydrodynamic simulations (Yoshikawa et al.~2001; White,
Hernquist \& Springel 2002).  Combined with models of the halo mass
function and spatial clustering of halos, for which very accurate
fitting functions are obtained from $N$-body simulations (Jing 1998;
Sheth \& Tormen 1999; Hamana et al.~2001; Jenkins et al.~2001), the halo
approach predicts the spatial clustering of galaxies as well as their
number density (Mo \& Fukugita 1996; Mo, Mao \& White 1999).  
Comparing these predictions with the observed values, we
obtain constraints on the relation between galaxies and their hosting
halo mass for different populations of galaxies.
The latter methodology was first attempted by Jing \& Suto (1998) for LBGs 
at $z\sim 3$ using their halo catalogs from $N$-body simulations.

The outline of this paper is as follows.  Section 2 describes models and
basic equations.  Section 3 summarizes observational data that are used
to put constraints on the model parameters.  In section 4, our results
of LBGs at $z\sim4$ and $\sim 5$ are presented and are compared with
previous results of $z\sim3$ LBGs (Moustakas \& Somerville 2001; Bullock
et al.~2002).  We show results of LAEz5, and discuss their implications
in section 5.  Finally, section 6 is devoted to summary and discussion.
In Appendix, we illustrate how the HOF parameters depend on the shape of
the two-point correlation function.

Throughout this paper, we adopt a flat $\Lambda$CDM (Cold Dark Matter)
cosmology with the matter density $\Omega_{\rm m}=0.3$, the cosmological
constant $\Omega_\Lambda=0.7$, the Hubble constant $H_0=100h$km/s/Mpc
with $h=0.7$, and the normalization of the matter power spectrum
$\sigma_8=0.9$.  We adopt the fitting function of the CDM power spectrum
of Bardeen et al.~(1986).

\section{Halo approach for galaxy clustering} 

The basic idea behind the halo model that we adopt below has a long
history (Neyman \& Scott 1952; Limber 1953; Peebles 1974, 1980;
McClelland \& Silk 1977; and other recent papers referred to in this
section). The model predictions have been significantly improved with
the recent accurate models for the mass function, the biasing and the
density profile of dark matter halos (Seljak 2000; Peacock \& Smith
2000; Ma \& Fry 2000).  This approach has been applied to various
problems in cosmological nonlinear clustering, galaxy clustering and
weak lensing correlation (e.g., Sheth \& Jain 1997; 
Jing, Mo \& B\"orner 1998; Komatsu \& Kitayama
1999; Cooray, Hu \& Miralda-Escude 2000; Cooray \&
Sheth 2002; Scoccimarro et al.~2001; Shu, Mao \& Mo 2001; 
Hamana, Yoshida \& Suto 2002; 
Berlind \& Weinberg 2002; Bullock, Wechsler \& Somerville 2002; 
Moustakas \& Somerville 2002; Takada \& Jain 2003; Takada \& Hamana 2003).  
In this section, we summarize several expressions
which are most relevant to the current analysis.  In particular, we
mainly follow the modeling of Berlind \& Weinberg (2002), Bullock et
al.~(2002) and Moustakas \& Somerville (2002) in which readers may find
further details.

We adopt a simple parametric form for the average number of a given
galaxy population as a function of the hosting halo mass:
\begin{equation}
\label{eq:Ng}
  N_{\rm g}(M)=
\begin{cases}
{(M/M_1)^\alpha & ($M>M_{\rm min}$)  \cr 
0 & ($M<M_{\rm min}$) } 
\end{cases} .
\end{equation}
The above statistical and empirical relation is the essential ingredient
in the current modeling characterized by the minimum mass of halos which
host the population of galaxies ($M_{\rm min}$), a normalization parameter
which can be interpreted as the critical mass  above which halos typically
host more than one galaxy ($M_1$; note that $M_1$ may
exceed $M_{\rm min}$ since the above relation represents the statistically
expected value of the number of galaxies), and the power-law index of the
mass dependence of the efficiency of galaxy formation ($\alpha$).  We
will put constraints on the three parameters from the observed number
density and clustering amplitude for each galaxy population.  In short,
the number density of galaxies is most sensitive to $M_1$ which changes
the average number of galaxies per halo.  The clustering amplitude on
large angular scales ($>1\arcmin$) is determined by the hosting halos and thus
very sensitive to the mass of those halos, $M_{\rm min}$.  The
clustering on smaller scales, on the other hand, depends on those three
parameters in a fairly complicated fashion; roughly speaking, $M_{\rm
min}$ changes the amplitude, $\alpha$, and to a lesser extent $M_1$ as
well, changes the slope (see Appendix). 
Further detailed discussion may be found in
Berlind \& Weinberg (2002), Bullock et al.~(2002) and Moustakas \&
Somerville (2002).

With the above relation, the number density of the corresponding galaxy
population at redshift $z$ is given by
\begin{equation}
\label{eq:ng}
n_{g,z}(z)=\int_{M_{\rm min}}^{\infty} dM~n_{\rm halo}(M,z)~N_{\rm g}(M),
\end{equation}
where $n_{\rm halo}(M)$ denotes the halo mass function for which we
adopt the fitting function of Sheth \& Tormen (1999).

The galaxy two-point correlation function on small scales is dominated
by contributions of galaxy pairs located in the same halo.  We adopt the
following model (Bullock et al. 2002) for the mean number of galaxy
pairs $\langle N_{\rm g}(N_{\rm g}-1)\rangle(M)$ within a halo of mass
$M$:
\begin{equation}
\label{eq:ngpair}
\langle N_{\rm g}(N_{\rm g}-1)\rangle(M) =
\left\{
\begin{array}{l}
N_{\rm g}^2(M) \quad\quad\quad\quad \mbox{if } N_{\rm g}(M)>1\\
N_{\rm g}^2(M)\log(4 N_{\rm g}(M))/\log(4)\\
\quad\quad\quad\quad\quad \mbox{if } 1>N_{\rm g}(M)>0.25\\
0 \quad\quad\quad\quad\quad\quad\quad\quad\quad \mbox{otherwise.}
\end{array}
\right.
\end{equation}
The above empirical model is motivated by previous results from the
semi-analytic galaxy formation models (Benson et al.~2000; Kauffmann et
al.~1999) which indicate that for $N_{\rm g}(M)>1$ the scatter around
the mean number of galaxies is Poissonian, while for $N_{\rm g}(M)<1$ it
becomes sub-Poissonian.

In the framework of the halo model, the galaxy power spectrum consists
of two contributions, one from galaxy pairs located in the same halo
(1-halo term) and the other from galaxy pairs located in two different
halos (2-halo term):
\begin{eqnarray}
P_{\rm g}(k) = P_{\rm g}^{1h}(k)+ P_{\rm g}^{2h}(k) .
\end{eqnarray}
Assuming the linear halo bias model (Mo \& White 1996), the 2-halo term
reduces to
\begin{eqnarray}
\label{eq:P2h}
P_{\rm g}^{2h}(k) &=& P_{\rm lin}(k)\nonumber\\
&\times& 
\left[
\frac{1}{n_{g,z}}
\int dM~n_{\rm halo}(M) N_{\rm g}(M) b(M) y(k,M) 
\right]^2,
\end{eqnarray}
where $P_{\rm lin}(k)$ is the linear dark matter power spectrum, $b(M)$
is the halo bias factor (we adopt the modified fitting function of Sheth
\& Tormen 1999), and $y(k,M)$ is the Fourier transform of the halo dark
matter profile normalized by its mass, $y(k,M)=\tilde{\rho}(k,M)/M$.
See, e.g., section 3 of Seljak (2000) for details.  Here we assume that
galaxies in halos trace the density profile of the underlying dark halos
by Navarro, Frenk \& White (1996; 1997), and adopt the
mass-concentration parameter relation by Bullock et al.~(2001) but with
an appropriate correction (see Shimizu et al. 2003).  Since the
clustering on large scales is dominated by the 2-halo term, it is fairly
insensitive to the assumption of galaxy distribution inside the hosting
halo (Berlind \& Weinberg 2002).  It should be noted that since $y\simeq
1$ on large scales (e.g., scales much larger than the virial radius of
halos), on such scales the 2-halo term can be rewritten by
\begin{equation}
\label{eq:P2h-largescale}
P_{\rm g}^{2h}(k) = b_g^2(>M_{min}) P_{\rm lin}(k) 
\end{equation}
where the galaxy number weighted bias factor is defined by 
\begin{equation}
\label{eq:bias}
b_g(>M_{min})\equiv 
\frac
{\int dM~n_{\rm halo}(M) N_{\rm g}(M) b(M)}
{\int dM~n_{\rm halo}(M) N_{\rm g}(M)}.
\end{equation}
Note that $b_g(>M_{min})$ depends on $\alpha$.

The 1-halo term is written as
\begin{eqnarray}
\label{eq:P1h}
P_{\rm g}^{1h}(k)&=&{1 \over {(2 \pi)^3 n_{g,z}^2}} 
\int dM~ n_{\rm halo}(M)~ \langle N_{\rm g}(N_{\rm g}-1) \rangle(M) \cr
&\times& |y(k,M)|^p .
\end{eqnarray}
We choose $p=2$ for $\langle N_{\rm g}(N_{\rm g}-1) \rangle >1$ and
$p=1$ for $\langle N_{\rm g}(N_{\rm g}-1) \rangle <1$ (Seljak 2000).
Here we assume that in the limit of a small number of galaxies in one
halo, one galaxy is located near the center of the halo.  Therefore, in
this case, the number of pairs is dominated by the central galaxy paired
with a halo galaxy, and thus the probability of finding a galaxy pair is
given by the single density profile of the galaxies within a halo.

Once the power spectrum of the galaxy population is specified, one can
easily compute their angular two-point correlation function via the
Limber projection (see e.g, chapter 2 of Bartelmann \& Schneider 2001):
\begin{equation}
\label{eq:limbers}
\omega(\theta) = \int dr~ q^2(r) \int {{dk}\over {2 \pi}}
k~ P_{\rm g} (k,r)~ J_0[f_K(r)\theta k],
\end{equation}
where $q(r)$ is the normalized selection function and $J_0(x)$ is the
zeroth-order Bessel function of the first kind.  For the spatially flat
cosmology ($\Omega_{\rm m}+\Omega_\Lambda=1$) as we consider throughout
the present paper, the radial function $f_K(r)$ is equivalent to $r$,
and $r=r(z)$ is the radial comoving distance given by
\begin{equation}
\label{eq:r-z}
r(z) = \frac{c}{H_0} \int_0^z
\frac{dz}
{\sqrt{\Omega_{\rm m}(1+z)^3 + \Omega_\Lambda}} .
\end{equation}
The dependences of HOF parameters on the shape of the angular two-point
correlation function are summarized in Appendix.

For a given selection function of the observation, the average galaxy
number density is
\begin{equation}
\label{eq:nga}
n_{\rm g} = \frac{\displaystyle \int dz~ \frac{dV(r)}{dz} q(r) n_{g,z}(r)}
{\displaystyle \int dz~ \frac{dV(r)}{dz} q(r)},
\end{equation}
where $dV(r)/dz$ denotes the comoving volume element per unit solid
angle:
\begin{eqnarray}
\frac{dV}{dz} = r^2(z) \frac{dr}{dz}
= \frac{c}{H_0} \frac{r^2}{\sqrt{\Omega_{\rm m} (1 + z)^3 +\Omega_\Lambda }} ,
\end{eqnarray}
again for the spatially flat cosmology.

\begin{figure}
\begin{center}
\begin{minipage}{8.4cm}
\epsfxsize=8.4cm 
\epsffile{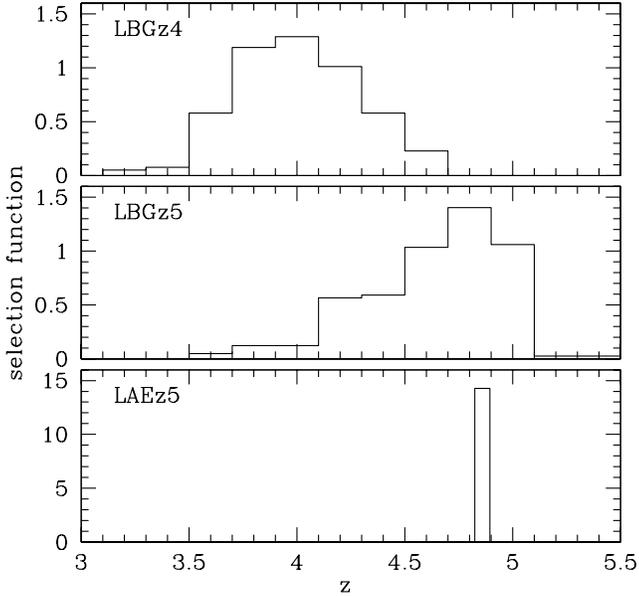}
\end{minipage}
\end{center}
\caption{Selection functions as a function of redshift for LBGz4 (top),
LBGz5 (middle) and LAEz5 (bottom).}  \label{fig:selection}
\end{figure}

\begin{figure}
\begin{center}
\begin{minipage}{8.4cm}
\epsfxsize=8.4cm 
\epsffile{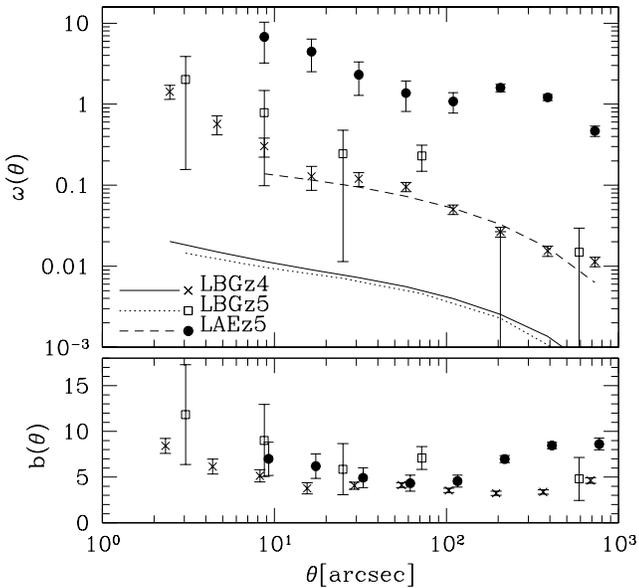}
\end{minipage}
\end{center}
\caption{Angular two-point correlation functions (upper panel) and the
corresponding bias parameter (lower panel) for three populations of
high-redshift galaxies.  crosses, open squires, and filled
circles represent the measured angular two-point correlation functions,
$w_{\rm g}(\theta)$, for LBGz4, LBGz5, and LAEz5, respectively.  In the
lower panel, points for LBGz4 and LAEz5 are slightly shifted
horizontally for clarify.  The measured correlation functions plotted
have been corrected for contamination and for integration constant (see
text).  The model predictions for dark matter angular two-point
correlation functions, $w_{\rm dm}(\theta)$, in the $\Lambda$CDM
cosmology with the same selection functions are plotted in curves; LBGz4
(solid line), LBGz5 (dotted line) and LAEz5 (dashed line).  To compute
them, the nonlinear fitting function of the CDM power spectrum by
Peacock \& Dodds (1996) is used.  The biasing parameter is simply
defined as $b(\theta)\equiv \sqrt{w_{\rm g}(\theta)/w_{\rm
dm}(\theta)}$.}  \label{fig:angcor}
\end{figure}

\section{Data}

\begin{table}
\caption{Summary of observed properties, here the large scale bias is
defined by $b=\sqrt{w_g/w_{dm}}$ on scales $\theta>100\arcsec$.}
\label{table:data}
\begin{tabular}{lcc} 
\hline
Sample & number density [$h^{3}$Mpc$^{-3}$] & large scale bias \\
\hline
LBGz4 & $(5.86\pm 0.71)\times 10^{-3}$ & $3 - 4.5$\\
LBGz5 & $(8.05\pm 4.96)\times 10^{-4}$ & $5 - 7$\\
LAEz5 & $(3.01\pm 1.94)\times 10^{-3}$ & $5 - 9$\\
\hline
\end{tabular}
\end{table}

We use three different samples of galaxy populations from deep imaging
data taken as part of the Subaru Deep Field (SDF) survey; LBGs at $z\sim
4$ (LBGz4s), LBGs at $z\sim 5$ (LBGz5s) and LAEs at $z\simeq 4.86$
(LAEz5s).  Observational details of those samples are described in Ouchi
et al.~(2001, 2003a, 2003b), and thus we briefly summarize their basic
features relevant for our comparison.

LBGz4s are selected from an $i'$-detection catalog constructed from deep
$BRi'$ imaging data over a 543 arcmin$^2$ area in the SDF (Ouchi et
al.~2003a).  The limiting AB magnitudes for the 3 $\sigma$ detection of
an object in a 1\arcsec.8 diameter aperture are $B=27.8$, $R=27.1$ and
$i'=26.9$.  In order to guarantee a reasonable level of photometric
completeness, the $i'$-detection catalog is limited to $i'=26.5$,
corresponding to the absolute magnitude of $M_{\rm 1700}=-19.0+5\log h$
for LBGz4s.  A total of 1438 LBGz4 candidates are detected in a range of
$3.5 < z < 4.5$.  The selection function is shown in top-panel of Figure
1, which is estimated using the Monte-Carlo simulation on the basis of
colors and redshifts of Hubble Deep Field North galaxies given in the
photometric redshift catalog by Furusawa et al.~(2000; see 
Ouchi et al.~2003c, Ouchi 2003 for details).  

It should be noted that the selection functions plotted in Figure 1 are
the probability distribution functions of galaxy redshifts in our
samples.  The current observational method does not specify the redshift
of individual galaxies accurately except in a statistical sense.  Their
number density is estimated to be $n_{\rm LBGz4}= (5.86 \pm 0.71)\times
10^{-3}h^{3}$Mpc$^{-3}$.  The angular two-point correlation function is
computed by the procedure described in Ouchi et al.~(2001; 2003d) and is
plotted in Figure \ref{fig:angcor}.

LBGz5s are selected from a $z'$-detection catalog constructed from deep
$Vi'z'$ imaging data over a 616 arcmin$^2$ area in the SDF (Ouchi et
al.~2003b).  The limiting AB magnitudes are $B=27.8$, $V=27.3$,
$R=27.1$, $i'=26.9$, $z'=26.1$ for the 3 $\sigma$ detection in a
1\arcsec.8 diameter aperture.  Again for the photometric completeness,
the $z'$-detection catalog is limited to $z'=26.0$, corresponding to the
absolute magnitude of $M_{\rm 1700}=-19.7+5\log h$ for LBGz5s.  A total
of 246 LBGz5 candidates are detected in a range of $4.2<z< 5.2$.  Their
number density is estimated to be $n_{\rm LBGz5}=(8.05 \pm 4.96) \times
10^{-4}h^{3}$Mpc$^{-3}$.

\begin{figure*}
\begin{center}
\begin{minipage}{13.4cm}
\epsfxsize=13.4cm 
\epsffile{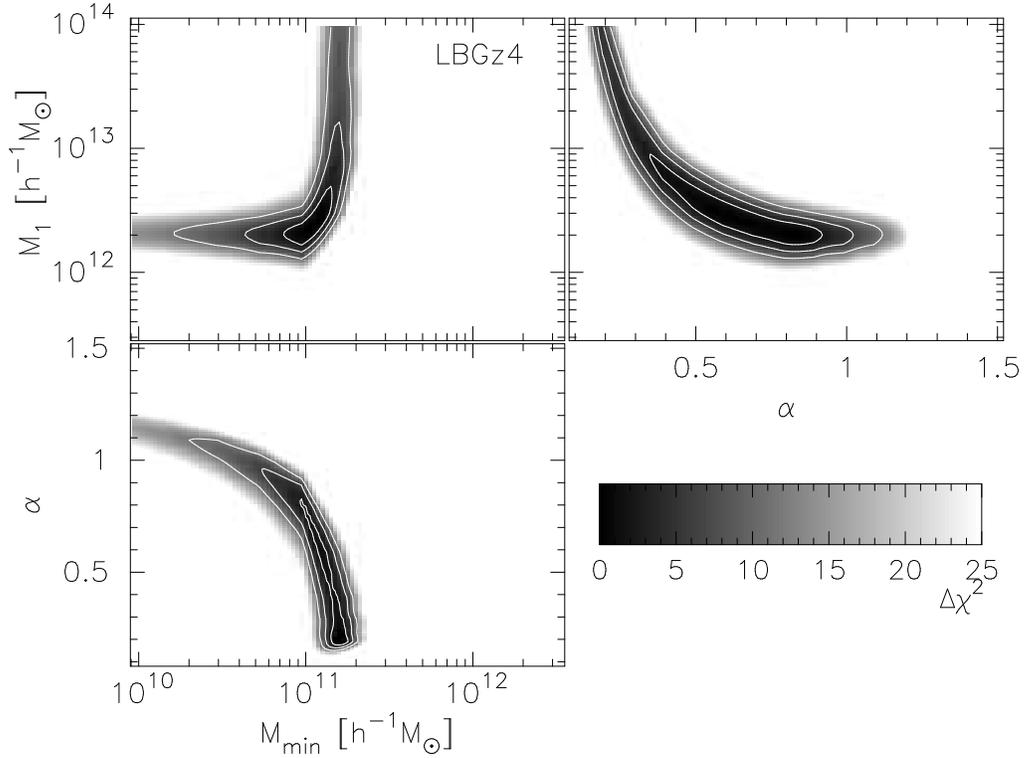}
\end{minipage}
\end{center}
\caption{Confidence contour maps derived from $\Delta \chi^2$ for LBGz4
on the two-parameter plane after marginalizing over the remaining one
parameter.  Top-left panel is on $M_{\rm min}$-$M_1$, top-right on
$\alpha$-$M_1$ and bottom-left on $M_{\rm min}$-$\alpha$.  A darker
gray-scale indicates a lower $\Delta \chi^2$ value (thus more likely).
Contour lines indicate from inner to outer $\Delta\chi^2=2.3$, 6.17 and
11.8, which, if each bin of the correlation function is independent,  
correspond to 68.3, 95.4, and 99.73\% confidence levels, respectively. 
In the present analysis, these confidence levels should be understood 
as approximate estimates.}
\label{fig:Lmap-LBGz4}
\end{figure*}

\begin{figure*}
\begin{center}
\begin{minipage}{13.4cm}
\epsfxsize=13.4cm 
\epsffile{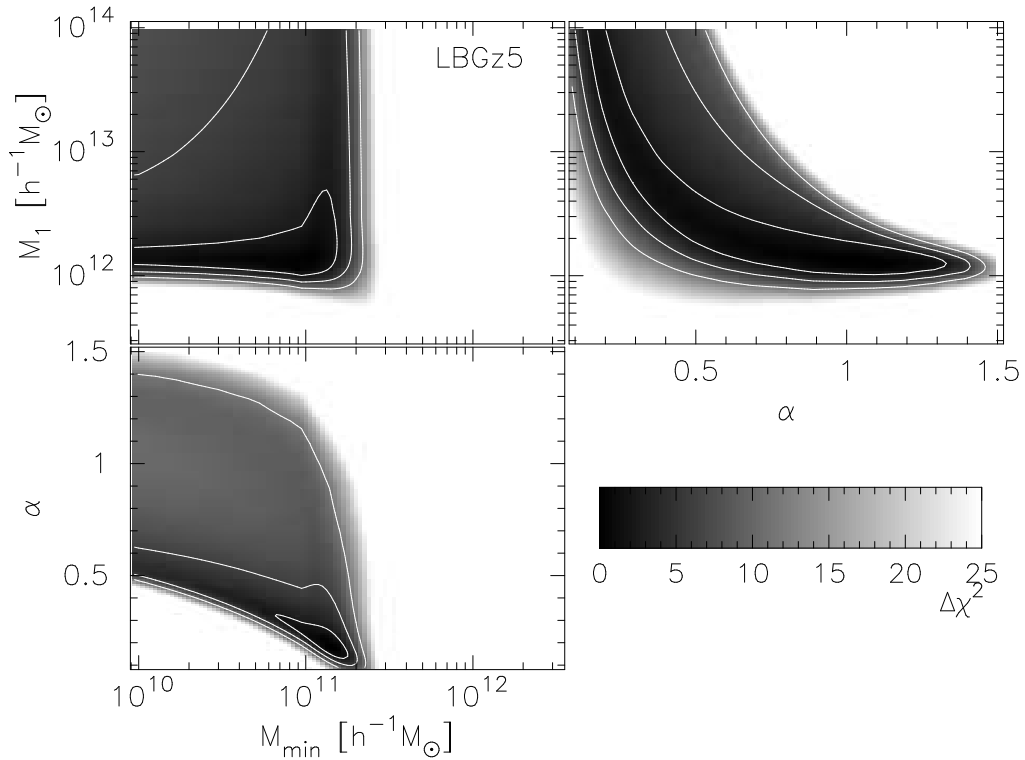}
\end{minipage}
\end{center}
\caption{Same as Figure \ref{fig:Lmap-LBGz4} but for LBGz5.}
\label{fig:Lmap-LBGz5}
\end{figure*}

LAEz5s were first selected from the same data of LBGz5s but an additional
observation using a narrow band-filter (NB711, central wavelength of
$7126\pm4$ \AA, FWHM bandwidth of $73.0\pm0.6$ \AA) was performed to
identify LAEs at $z\simeq 4.86$ (Ouchi et al.~2003a).  The limiting
magnitude is NB711$=26.0$ for the 3 $\sigma$ detection in a 1\arcsec.8
diameter aperture.  The selection function of LAEz5 is approximated by a
top-hat function (bottom-panel of Figure 1) whose shape (center and
width) is determined from the central wavelength and FWHM of the NB711
filter (Ouchi et al. 2003a).  A total of 87 LAEz5 candidates are
detected over $4.83 \la z \la 4.89$.  Their number density is $n_{\rm
LAEz5}=(3.01 \pm 1.94)\times 10^{-3}h^{3}$Mpc$^{-3}$.

The angular two-point correlation function $w_{\rm g}(\theta)$ for each
sample is plotted in Figure \ref{fig:angcor} together with that for dark
matter, $w_{\rm dm}(\theta)$, which is calculated applying the same
selection function of the observational data.  The plotted correlation
functions have been corrected for contamination and for the ``integration
constant'' (see Groth \& Peebles 1977).  Possible
contaminants are mainly low-$z$ galaxies which match the selection
criteria by chance, and thus we may safely assume no cross-correlation 
between high-$z$ LBGs/LAEs and low-$z$ contaminants and no 
auto-correlation in low-$z$ contaminants. Therefore we have made
contamination correction for each sample by multiplying the observed
correlation function by a factor of $1/(1-f_c)^2$, where the
contamination rate $f_c$ is estimated to be 0.01, 0.26 and 0.4, for
LBGz4s, LBGz5s and LAEz5s, respectively.  The integration constant is
estimated to be 0.00676, 0.00637 and 0.00675 for LBGz4, LBGz5 and LAEz5,
respectively (Ouchi et al.~2001, 2003a, 2003b).  We define the biasing
parameter as $b(\theta)\equiv \sqrt{w_{\rm g}(\theta)/w_{\rm
dm}(\theta)}$.  For LBGs, we obtain large-scale (specifically, $60
\arcsec <\theta < 1000 \arcsec$) bias factors of $b=3 - 4.5$ and
$5 - 7$ for LBGz4 and LBGz5s, respectively, and the bias increases
with decreasing the separation.  On the other hand, LAEs exhibit
stronger clustering on larger scales ($\theta >200 \arcsec$) than LBGs,
while their biases on smaller scales are similar to those of LBGs.

\section{Results on Lyman-break galaxies}

\subsection{Constraints on $M_{\rm min}$, $M_1$, and $\alpha$}

We estimate the range of allowed values for the three parameters,
$M_{\rm min}$, $M_1$, and $\alpha$ by considering the following $\chi^2$
function constructed from the observed number density and the angular
two-point correlation functions:
\begin{eqnarray}
\label{eq:chi2}
\chi^2(M_{\rm min},M_1,\alpha) &=& 
\sum_{\theta_{\rm bin}} 
{{[\omega^{\rm obs}(\theta_{\rm bin})-\omega^{\rm model}(\theta_{\rm bin})]^2}
\over {\sigma_{\omega}^2(\theta_{\rm bin})}}\nonumber\\
&&+{{[\log n_{\rm g}^{\rm obs}-\log n_{\rm g}^{\rm model}]^2} \over {\sigma_{\log n_{\rm g}}^2}},
\end{eqnarray}
where $\sigma_{\omega}$ and $\sigma_{\log n_{\rm g}}$ are the
statistical 1-$\sigma$ error in the measurements of the angular
correlation function and the number density, respectively.  In the above
likelihood estimator, we take the logarithm of the galaxy number density
instead of the number density itself, because the predicted galaxy
number density varies logarithmically with $M_1$.  Note that although
the HOF parameters can depend on time in general, we assume here that
the three parameters are constant within the redshift interval of each
sample for simplicity.  For LAEz5s, this must be the case as the
redshift interval is very small. It turns out (see \S \ref{LBGevo}) that
the HOF parameters for LBGs do not change significantly over three LBG
samples at $z\sim3$, $z\sim4$ and $z\sim5$.  Therefore the above
assumption is reasonable for LBGs as well.  We also note that the
present analysis does not take the cosmic variance into account although
it may be important for relatively small survey volumes for those
samples.  Figures \ref{fig:Lmap-LBGz4} and \ref{fig:Lmap-LBGz5} show the
$\chi^2$ map on two-parameter planes after marginalizing over the
remaining one parameter. The two-dimensional likelihood contours
represent $\Delta\chi^2=2.3$, 6.17 and 11.8 which, if each bin of the
correlation function is independent, should correspond to 68.3, 95.4 and
99.7\% confidence levels (Press et al.~1986). Strictly speaking,
however, the sampled correlation function bins are not completely
independent, and these confidence levels should be regarded simply as
approximate estimates.

\begin{figure}
\begin{center}
\begin{minipage}{8.4cm}
\epsfxsize=8.4cm 
\epsffile{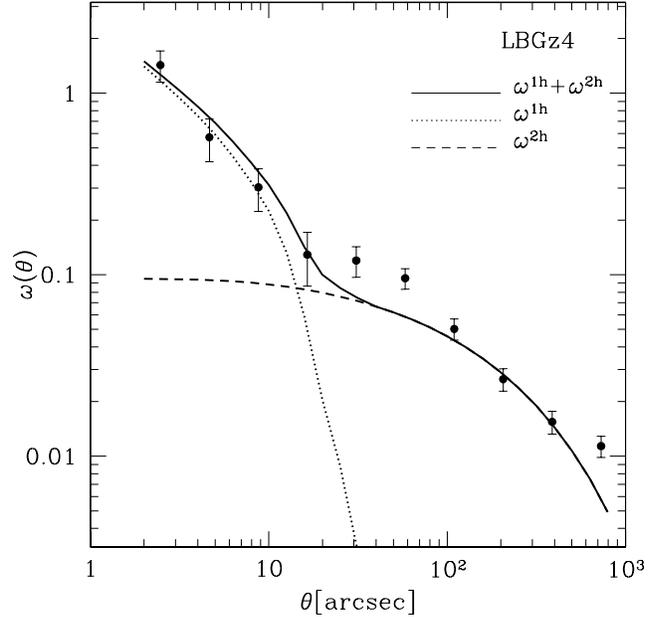}
\end{minipage}
\end{center}
\caption{Comparison of the observed angular correlation function of
LBGz4s with the model prediction assuming 
$M_{\rm min}=1.6\times 10^{11}h^{-1}M_\odot$,
$M_1=8\times 10^{12}h^{-1}M_\odot$, and $\alpha=0.75$.}
\label{fig:angcor_LBGz4}
\end{figure}

\begin{figure}
\begin{center}
\begin{minipage}{8.4cm}
\epsfxsize=8.4cm 
\epsffile{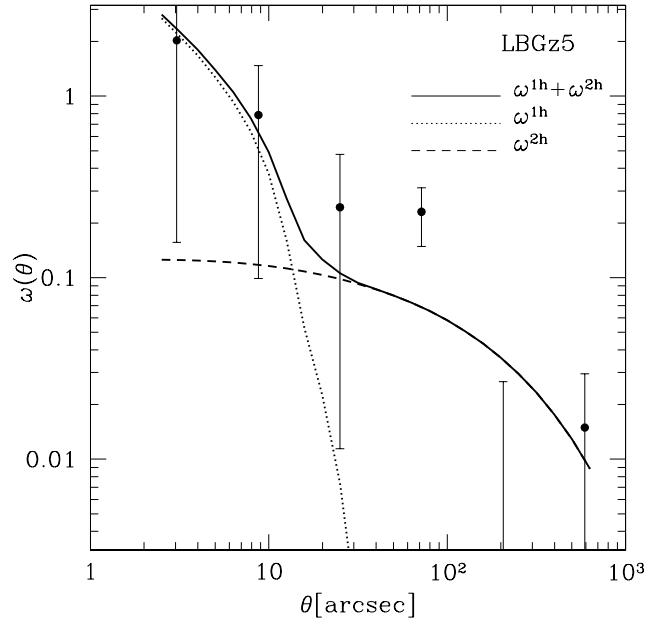}
\end{minipage}
\end{center}
\caption{Same as Figure \ref{fig:angcor_LBGz4} but for LBGz5s assuming
$M_{\rm min}=1.5\times 10^{11}h^{-1}M_\odot$, $M_1=5\times
10^{12}h^{-1}M_\odot$, and $\alpha=0.75$.}
\label{fig:angcor_LBGz5}
\end{figure}

Examine first the parameters for LBGz4s which are fairly strongly
constrained by the observations (Fig. \ref{fig:Lmap-LBGz4}).  Top-left
panel shows the likelihood map on $M_{\rm min}$-$M_1$ plane.  As pointed
out earlier by Berlind \& Weinberg (2002), Bullock et al. (2002), and
Moustakas \& Somerville (2002), $M_1$ and $M_{\rm min}$ are mainly
constrained by the number density and their clustering amplitude on
large scales ($\theta>1\arcmin$), respectively.  As Figure
\ref{fig:angcor} shows, the observational uncertainty in the clustering
amplitude for LBGz4s is fairly small and $\delta n/n \sim 12$\%.  Thus
we have relatively tight constraints on those two parameters.  The
constraint on $\alpha$ is weak because of the degeneracy with the other
two parameters.  However, it is clear that the data favor $\alpha<1$
implying that the galaxy formation is less efficient (or small galaxies
merge more efficiently to form larger ones) in more massive halos.

Turn next to LBGz5 (Fig. \ref{fig:Lmap-LBGz5}). The constraints on this
population are not so tight because of much larger uncertainties in the
clustering amplitude (Fig. \ref{fig:angcor}) and in the number density,
$\delta n/n \sim 62$\%.  Nevertheless the constraints on the parameters
for LBGz5s seem very similar to those for LBGz4, and we are not able to
detect any significant difference of the parameter values of LBGs
between $z=4$ and 5.

Figures \ref{fig:angcor_LBGz4} and \ref{fig:angcor_LBGz5} compare the
observed angular two-point correlations with the halo model predictions
based on preferred parameters for LBGz4s and LBGz5, respectively.  In
plotting the model predictions, we adopt $M_{\rm min}=1.6\times
10^{11}h^{-1}M_\odot$, $M_1=8\times 10^{12}h^{-1}M_\odot$, and
$\alpha=0.75$ for LBGz4, and $M_{\rm min}=1.5\times
10^{11}h^{-1}M_\odot$, $M_1=5\times 10^{12}h^{-1}M_\odot$, and
$\alpha=0.75$ for LBGz5.  Given the approximate and empirical nature of
the halo model, the overall agreement is satisfactory.

\subsection{Characteristics of LBG hosting halos}

\begin{figure}
\begin{center}
\begin{minipage}{8.4cm}
\epsfxsize=8.4cm 
\epsffile{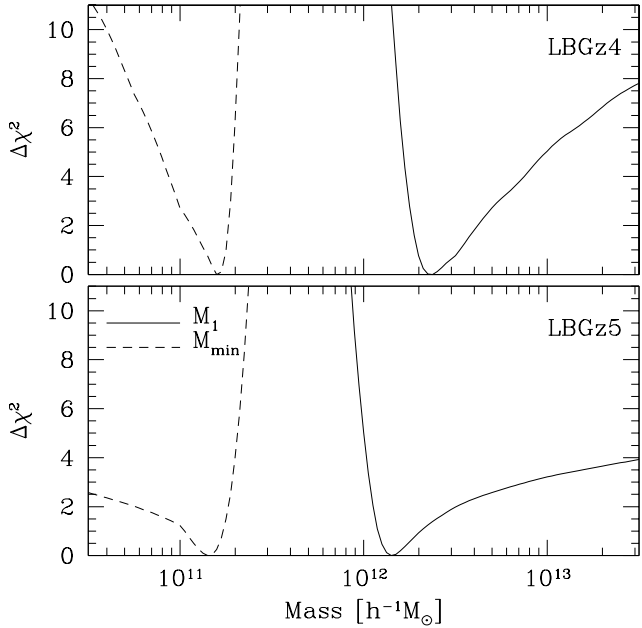}
\end{minipage}
\end{center}
\caption{The likelihood functions ($\Delta\chi^2$) for $M_{\rm min}$
(dashed lines) and for $M_1$ (solid lines) obtained after marginalizing
over the other two parameters.  Upper and lower panel is for LBGz4 and 
LBGz5, respectively.}  \label{fig:likelymass}
\end{figure}

Let us look into more carefully the hosting halo masses for the two LBG
samples.  Figure \ref{fig:likelymass} plots the likelihood functions
($\Delta \chi^2$) for $M_{\rm min}$ (dashed lines) and for $M_1$ (solid
lines) after marginalizing over the remaining two parameters.  Clearly,
both likelihood functions look similar, indicating little evolution of
properties of the hosting halos from $z\sim 5$ to $z\sim 4$.

In order to estimate the characteristic mass of hosting halos and the
typical number of galaxies per halo, we introduce the following two
quantities: the average mass of the hosting halo (member galaxy number
weighted):
\begin{equation}
\label{eq:average-mass}
\langle M_{\rm host} \rangle = 
\frac{\int_{M_{\rm min}}^\infty dM~ M~ N_{\rm g}(M) n_{\rm halo} (M)}
{\int_{M_{\rm min}}^\infty dM~ N_{\rm g}(M) n_{\rm halo} (M)},
\end{equation}
and the average number of galaxies per halo:
\begin{eqnarray}
\label{eq:average-number}
 \langle N_{\rm g} \rangle \equiv 
\frac{\int_{M_{\rm min}}^\infty N_{\rm g}(M) \, n_{\rm halo}(M,z)dM} 
{\int_{M_{\rm min}}^\infty  n_{\rm halo}(M,z)dM} .
\end{eqnarray}
Those are evaluated assuming typical sets of HOF parameters that we
found in the previous subsection (Table \ref{table:average}).  The
average mass of the hosting halos for LBGs is $(5 - 6)\times
10^{11}h^{-1}M_\odot$, and the average number of galaxies per halo is
$\sim 0.4$.  Thus LBGs have an approximate one-to-one correspondence to
relatively less massive halos.

\begin{table}
\caption{The galaxy-number weighted average mass of hosting halo
$\langle M_{\rm host} \rangle$  (in units of $h^{-1}M_\odot$) and the 
expected number of galaxies per one halo $\langle N_{\rm g} \rangle$ 
for typical values of HOF parameters.} \label{table:average}
\begin{tabular}{lcc} 
\hline
Sample ($M_{\rm min}$, $M_1$, $\alpha$) 
& $\langle N_{\rm g} \rangle$ & $\langle M_{\rm halo} \rangle$ \\
\hline
LBGz4 ($1.6\times10^{11}$, $2.4\times 10^{12}$, $0.5$) & 
0.38  & $6.3\times 10^{11}$ \\
LBGz5 ($1.4\times10^{11}$, $1.4\times 10^{12}$, $0.5$) & 
0.45  & $4.5\times 10^{11}$ \\
\hline
\end{tabular}
\end{table}

\begin{figure*}
\begin{center}
\begin{minipage}{16cm}
\epsfxsize=16cm 
\epsffile{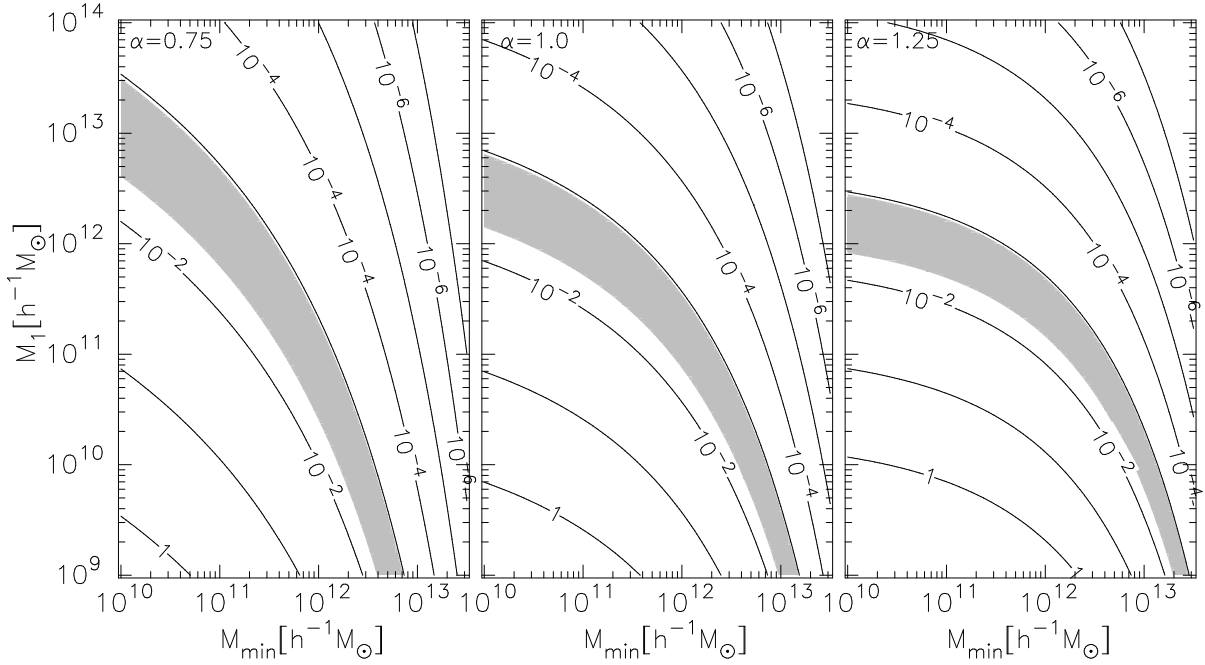}
\end{minipage}
\end{center}
\caption{Contours show the halo model prediction for the number 
density of LAEz5s. Grayed region shows the 1-$\sigma$ range of 
the observed value of $n_g=(3.01\pm1.94)\times10^{-3}(h \mbox{Mpc}^{-1})^3$. 
{}Form left to right $\alpha=1.25$, 1.0 and 0.75, respectively.}
\label{fig:LAEdensity}
\end{figure*}

\subsection{Evolution of properties of the hosting halos for LBGs}
\label{LBGevo}

Turn next to the evolution of the hosting halos for LBGs. Our current
analysis indicates that the minimum halo mass $M_{\rm min}$ for LBGs is
almost the same $\sim 1.5\times 10^{11}h^{-1}M_\odot$ at $z\sim4$
and $z\sim 5$. The halo mass accommodating more than one galaxy, $M_1$,
seems increasing as time although barely at a 1-$\sigma$ level.  

The almost identical analyses by Moustakas \& Somerville (2001) and
Bullock et al.~(2002) for $z\sim3$ LBGs (Steidel et al. 1998; Adelberger
et al. 1998; Adelberger 2000) indicate that $M_{\rm
min}=1.3\times10^{10}h^{-1}M_\odot$ and
$M_1=6\times10^{12}h^{-1}M_\odot$ for $\alpha=0.8$, and that $M_{\rm
min}=(0.4-8)\times10^{10}h^{-1}M_\odot$,
$M_1=(6-10)\times10^{12}h^{-1}M_\odot$ and $0.9<\alpha<1.1$,
respectively.  Taking account of the relatively large uncertainties in
those estimates, their results are consistent with ours, and indeed the
combined results may indicate an evolutionary trend of decreasing
$M_{\rm min}$ and increasing $M_1$ with decreasing $z$.  The different
selection criteria at different redshifts may induce an artificial
systematic effect in estimating hosting halo mass, but this is not the
case here.  If the limiting flux of a sample is brighter, galaxies in
the sample have a smaller number density and usually a higher clustering
amplitude.  This leads to increasing $M_{\rm min}$ and $M_1$
simultaneously.  However, the limiting absolute magnitudes for LBGz3s,
LBGz4s and LBGz5s are $M_{\rm 1700} =-19.3+5\log h$, $M_{\rm
1700}=-19.0+5\log h$ and $M_{\rm 1700}=-19.7+5\log h$, respectively.
Therefore, it is very unlikely that the difference in the limiting
magnitude solely accounts for the systematic (although weak) trends in
$M_{\rm min}$ and $M_1$.

In summary, the hosting halos for LBGs are characterized as follows; (i)
$M_{\rm min}$ is about $\simeq 1.5\times10^{11}h^{-1}M_\odot$ both at
$z\sim 4$ and $z\sim 5$, while it decreases to about $M_{\rm
min}=(0.4-8)\times10^{10}h^{-1}M_\odot$ at $z\sim 3$.  (ii) $M_1$
increases with time, $M_1\simeq 1.4 \times10^{12}$, $2.4 \times10^{12}$
and $(6-10) \times10^{12}h^{-1}M_\odot$ for $z=5$, 4 and 3,
respectively.

\section{Results on Lyman-alpha emitters}

\begin{figure}
\begin{center}
\begin{minipage}{8.4cm}
\epsfxsize=8.4cm 
\epsffile{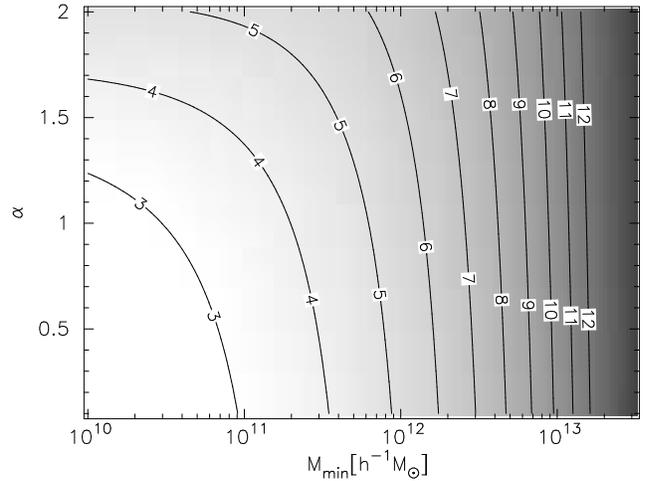}
\end{minipage}
\end{center}
\caption{Contours show the halo model prediction for the galaxy number
weighted large scale bias factor defined by eq.~(\ref{eq:P2h-largescale}).}
\label{fig:LAEbias}
\end{figure}

Let us turn to LAEz5s.  As shown in Figure \ref{fig:angcor}, their
angular correlation function exhibits a somewhat irregular shape.  This
is more clearly seen in the plot of the bias (lower panel of Figure
\ref{fig:angcor}).  On scales less than 120 arcsec, the bias increases
with decreasing separation similarly to LBGs, and its amplitude is in
the range between those for LBGz4s and LBGz5s.  On the other hand, on
larger scales the bias factor is rather high, which is in a marked
contrast with LBGs.  The number density of LAEs is higher than that of
LBGz5s but it has a large uncertainty, $\delta n/n \sim 65$\%.  It
should be noted that the survey volume of LAEz5 is small,
$(30h^{-1}\mbox{Mpc})^3$ (comoving volume), thus it is possible that
these measurements are significantly affected by the cosmic variance.

For reference, we give here some numbers which are useful in the
following discussion; at the redshift of LAEz5s, $z\simeq 4.86$, 1
arcmin corresponds to $1.56h^{-1}$Mpc (comoving), and the average number
of halos with the mass larger than $M$ within the survey volume computed
from the halo mass function is $N(>M)=90$, 10 and 1, for $M=3\times
10^{11}$, $1\times 10^{12}$, and $3\times 10^{12}h^{-1}M_\odot$,
respectively.

We apply the same likelihood analysis as performed for LBGs in the last
section, but we find that our simple HOF prescription fails to reproduce
simultaneously the observed angular correlation function and number
density.  Indeed, no parameter set is found to give a reasonably small
$\chi^2$.  The most serious discrepancy is the very high correlation
amplitude on scales larger than 120 arcsec.  In what follows, we present
two illustrative examples of failed models, which would help to search
for a possible solution of the problem.  For this purpose, we plot halo
model predictions for the number density of LAEz5s in Figure
\ref{fig:LAEdensity}, in which the gray region indicates the 1-$\sigma$
range of the observed number density, and halo model predictions for the
large scale bias factor defined by eq.~(\ref{eq:P2h-largescale}) in
Figure \ref{fig:LAEbias}.

\begin{figure}
\begin{center}
\begin{minipage}{8.4cm}
\epsfxsize=8.4cm 
\epsffile{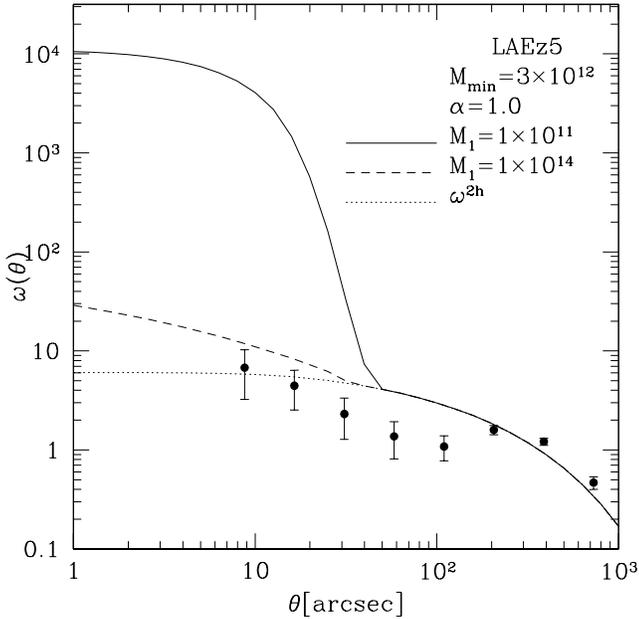}
\end{minipage}
\end{center}
\caption{Comparison of the observed angular correlation function of
LAEz5 with the model prediction assuming 
$\alpha=1.0$, $M_{\rm min}=3.0\times 10^{12}h^{-1}M_\odot$ 
and $M_1=1\times 10^{11}h^{-1}M_\odot$ for the solid line, 
and $M_1=1\times 10^{11}h^{-1}M_\odot$ for the dashed line.
The dotted line shows the 2-halo term only which does not 
depends on $M_1$.}
\label{fig:angcor_LAEz5-demo1}
\end{figure}

\begin{figure}
\begin{center}
\begin{minipage}{8.4cm}
\epsfxsize=8.4cm 
\epsffile{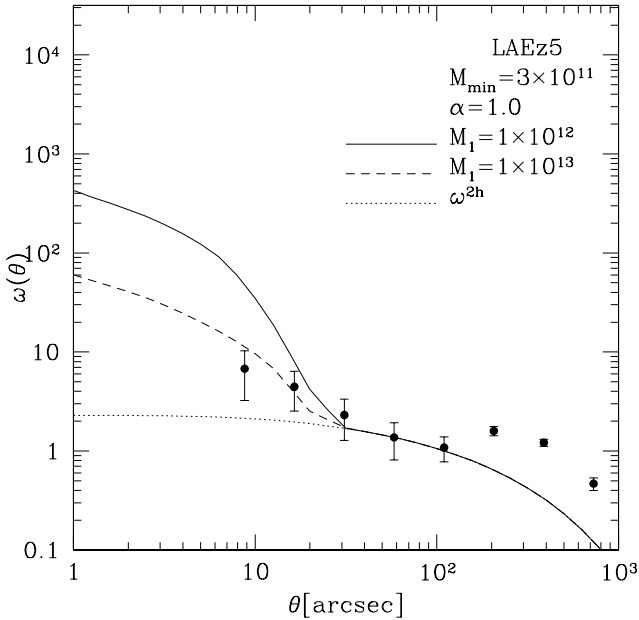}
\end{minipage}
\end{center}
\caption{Same as Figure \ref{fig:angcor_LAEz5-demo1} 
but for different HOF parameter
sets $M_{\rm min}=3.0\times 10^{11}h^{-1}M_\odot$, $\alpha=1.0$,
and $M_1=1.0\times 10^{12}$ (the solid line) and
$10^{13}$ (dashed line).}
\label{fig:angcor_LAEz5-demo2}
\end{figure}

The first example is as follows; taking the amplitude of the correlation
function on large scales ($b=7 - 9$ for $\theta>120$ arcsec, see Figure
\ref{fig:angcor}), one finds in Figure \ref{fig:LAEbias} that
$M_{min}=(2 - 5)\times 10^{12} h^{-1}M_\odot$ is required irrespective
of $\alpha$.  This value combined with the observed number density of
LAEz5s gives roughly $M_1\sim 10^{11}h^{-1}M_\odot$ (or less) for a
reasonable range of $\alpha$ (Fig. \ref{fig:LAEdensity}).  This
parameter set, however, predicts a much higher correlation amplitude
than observed on smaller scales (Fig. \ref{fig:angcor_LAEz5-demo1}).  If
one attempts to have a reasonable correlation amplitude on smaller
scale, a very large $M_1$ (more than $\sim10^{14}h^{-1}M_\odot$ at
least) is required (see Fig. \ref{fig:angcor_LAEz5-demo1}), which
leads to too small a number density on the other hand.

One possible way to reconcile this discrepancy is to modify the halo
model that we have adopted.  Figure \ref{fig:angcor_LAEz5-demo1}
indicates that it is the 1-halo term that boosts the correlation
amplitude on smaller scales.  Therefore, our assumption in the halo
model that the galaxy distribution follows the dark matter distribution
may not hold for LAEz5s.  In this paper, however, we do not attempt to
develop the HOF model by allowing a possible variation on the galaxy
distribution within halos, because the statistical accuracies of the
correlation functions on small scales are still relatively low.  Indeed,
the number of small separation pairs is very small; the number of LAE
pairs which fall into the smallest separation bin is two, and that into
the second bin is five.  This means that halos having more than one LAE
are limited, and one has to keep in mind that the correlation function
measurements on these scales are based on LAEs in such a small number of
halos.  Given a possible large uncertainty in the measurements of
small-scale clustering, it is premature to explore an alternative model
in great detail at this point.
We also note that given the limited number of LAEz5s, it is not to be 
denied that their correlation signals are contaminated by the presence 
of the possible foreground galaxies in one or a few clusters.
We must wait for future data extending over a much larger survey area.

In addition to the above problem, there is another problem concerning
the predicted number of halos in the survey volume.  If one takes
$M_{min} = 3\times 10^{12}h^{-1}M_\odot$, the expected mean number of
halos with mass larger than this value in the survey volume is
$N(>3\times 10^{12}h^{-1}M_\odot)\sim 1$.  This suggests that there are
only a few hosting halos in the survey volume.  If this picture is
right, it is unlikely that the 2-halo term is correctly measured from
such a small number of halos.

The above consideration implies that a simple halo model may not be
readily applicable to the LAEs.  Note that the parameter set ($M_{min}$,
$M_1$, $\alpha$)$=$($3\times 10^{12}h^{-1}M_\odot$, $1\times
10^{11}h^{-1}M_\odot$, $1.0$) gives $\langle N_g \rangle=45$ and
$\langle M_{\rm halo}\rangle =5.2\times 10^{12}h^{-1}M_\odot$.
Therefore, if real, we can conclude at least that the nature of hosting
halos and the relation between galaxies and halos for LAEs are very
different from those for LBGs.

The second example is as follows.  Let us ignore the data on large
scales ($\theta>120$ arcsec) for a while, and adopt the value of $b$
around 1 arcmin, i.e., $b \simeq 4$.  This gives $M_{min}=(2 -4)
\times 10^{11} h^{-1}M_\odot$ (Fig. \ref{fig:LAEbias}).  This value
combined with the observed number density gives roughly $M_1\sim
10^{12}h^{-1}M_\odot$ for a reasonable range of $\alpha$ (Fig.
\ref{fig:LAEdensity}).  This parameter set predicts a slightly higher
correlation amplitude on small scales than observed as shown in Figure
\ref{fig:angcor_LAEz5-demo2}.  However, this discrepancy may not be
taken so seriously, because of the limited statistical significance 
as mentioned above.  Better agreement is obtained by setting
larger $M_1\sim 10^{13}h^{-1}M_\odot$ (Fig.
\ref{fig:angcor_LAEz5-demo2}), which, however, predicts a smaller number
density than observed.  The discrepancy in the number density becomes
smaller if one takes a smaller $\alpha$ (Fig. \ref{fig:LAEdensity}).  If
we take ($M_{min}$, $M_1$)$=$($3\times 10^{11}h^{-1}M_\odot$, $1\times
10^{12}h^{-1}M_\odot$), the characteristic values are $\langle N_g
\rangle=0.67$ and $\langle M_{\rm halo}\rangle =9.0\times
10^{11}h^{-1}M_\odot$ for $\alpha=0.75$, and $\langle N_g \rangle=0.75$
and $\langle M_{\rm halo}\rangle =7.8\times 10^{11}h^{-1}M_\odot$ for
$\alpha=0.5$.  These values are similar to those for LBGs.

The above model provides acceptable agreement with both the correlation
function on scales smaller than 120 arcsec and the number density.  The
predicted correlation function on larger scales, however, has a much
lower amplitude than observed.  A possible interpretation to this
discrepancy is that the current survey volume is one of the overdense
regions on large scales by chance, and the LAEs in the region have
accidentally acquired the high correlation amplitude.  In fact,
Shimasaku et al. (2003) have extended the survey area of LAEz5s to the
north and found a high overdensity of LAEz5s over a circular region of 5
arcmin (8$h^{-1}$Mpc) radius.  They have suggested that it may be a
progenitor of a present-day massive cluster of galaxies.  Half of this
circular region is inside our survey area.  LAEz5s associated with this
large-scale overdense region will have an unusually high correlation
amplitude on large scales.  If this is the case, the HOF parameter
values obtained ignoring the large-scale correlation function will be
close to the typical values of LAEs at $z\sim 5$.

\section{Summary and discussions}

We have analyzed three high-redshift galaxy samples created from the
Subaru Deep Field (SDF) survey data; LBGs at $z\sim 4$ (LBGz4s), LBGs at
$z\sim 5$ (LBGz5s) and LAEs at $z\simeq 4.86$ (LAEz5s), and explored the
implications of their number density and angular clustering in the
framework of the halo occupation function (HOF).

Our major findings are summarized as follows;

(i) The two LBG samples can be well described by the halo model with an
appropriate HOF in an approximate fashion.

(ii) The hosting halos for LBGz4s and LBGz5s are more massive than $M_{\rm
min} \sim 1.5\times 10^{11}h^{-1}M_\odot$. Since the expected number of
LBGs per halo with $M>M_{\rm min}$ is $\sim 0.5$, there is an
approximate one-to-one correspondence between halos and LBGs. This is
basically consistent with the results previously found for LBGs at
$z\sim 3$ (Mo, Fukugita 1996; Steidel et al. 1998; Jing \& Suto 1998
Moustakas \& Somerville 2001; Berlind \& Weinberg 2002; Bullock et
al. 2002). 

(iii) On the other hand, this may also indicate that a large fraction of
dark halos do not host LBGs brighter than $M_{\rm 1700} \simeq -19$ mag.
Nevertheless this does not necessarily mean that there is no galaxy in
such halos.  Franx et al. (2003) have found a population of red galaxies
at $z\sim 3$ which do not have active star formation and so may not be
easily detectable by the Lyman break technique because of the very faint
UV continuum emission.  They estimate that the number density of such
red galaxies is about half that of LBGs at the same redshift.  Thus it
may be the case that a fraction of the dark halos at $z \sim 4-5$ host
such red galaxies rather than bright LBGs that we have discussed here.

(iv) The LBG samples at $z\sim 3$, 4 and 5 discussed here have very
similar limiting absolute magnitudes, $M_{\rm 1700}\simeq -19$.  On the
other hand, the minimum mass of their hosting halos seems decreasing
with time, although its statistical significance is not strong. If true,
this means that the star formation efficiency per unit dark matter mass,
$L_{\rm 1700}/M_{\rm halo}$, increases with time.  This increase may
suggest that cold gas gradually accumulates in LBGs, if the star
formation rate is simply proportional to the amount of cold gas.

(v) There is a weak indication for $M_1$ to increase slightly with time
in the LBG samples.  If this is indeed the case, it may be explained by
the mutual merging of LBGs.

(vi) For LAEz5s, our simple HOF prescription fails to reproduce
simultaneously the observed angular correlation function and number
density.  No parameter set gives a reasonably small $\chi^2$.  This is
mainly because the LAEz5s exhibit very strong clustering signal on
scales larger than 120 arcsec.

In fact, the nature of LAEz5s is still uncertain in the current result;
models which match both the high correlation amplitude on large
scales and the number density of LAEz5s predict much higher correlation
amplitude on small scales than observed.  A possible interpretation of
this discrepancy is that the distribution of LAEs within halos differs
from that of dark matter.  
If this is the case, the simple halo model description for the LAE we
adopted in this paper needs to be improved. 
Also the observational indication for the discrepancy should be carefully
confirmed with future larger and more accurate data samples.

Alternatively, if one constructs models which match both the correlation
function on small scales and the number density of LAEz5s, they imply a
lower correlation amplitude on large scales than observed.  This may be
simply due to statistical fluctuation in a sense that the current data
do not represent a fair sample of the universe at $z\sim 5$ as indicated
by by Shimasaku et al.~(2003).  The HOF parameters derived from the fit
to the observed data except for the large scale correlation function are
$M_{min}\sim 3\times 10^{11}h^{-1}M_\odot$ and $M_1\sim 1\times
10^{12}h^{-1}M_\odot$ for a reasonable range of $\alpha$.  This gives
$\langle N_g \rangle\sim0.7$ and $\langle M_{\rm halo}\rangle \sim
8\times 10^{11}h^{-1}M_\odot$ for $\alpha=0.75$, suggesting an
approximate one-to-one correspondence between LAEs and halos as found
for LBGs.

In order to distinguish the above two pictures, we need a much larger
observational sample after all.

\section*{Acknowledgments}
T.H. and M.O. acknowledge support from Japan Society for Promotion of
Science (JSPS) Research Fellowships.  I. K. gratefully acknowledges
support from the Takenaka-Ikueikai fellowship.  This research was also
supported in part by the Grants--in--Aid from Monbu--Kagakusho and Japan
Society of Promotion of Science (12640231, 13740150, 14102004, and
1470157).  Numerical computations presented in this paper were carried
out at ADAC (the Astronomical Data Analysis Center) of the National
Astronomical Observatory, Japan (project ID: mys02a, yys08a).

\appendix

\section{Dependences of HOF parameters on the shape of the two-point 
correlation function}

\begin{figure}
\begin{center}
\begin{minipage}{8.4cm}
\epsfxsize=8.4cm 
\epsffile{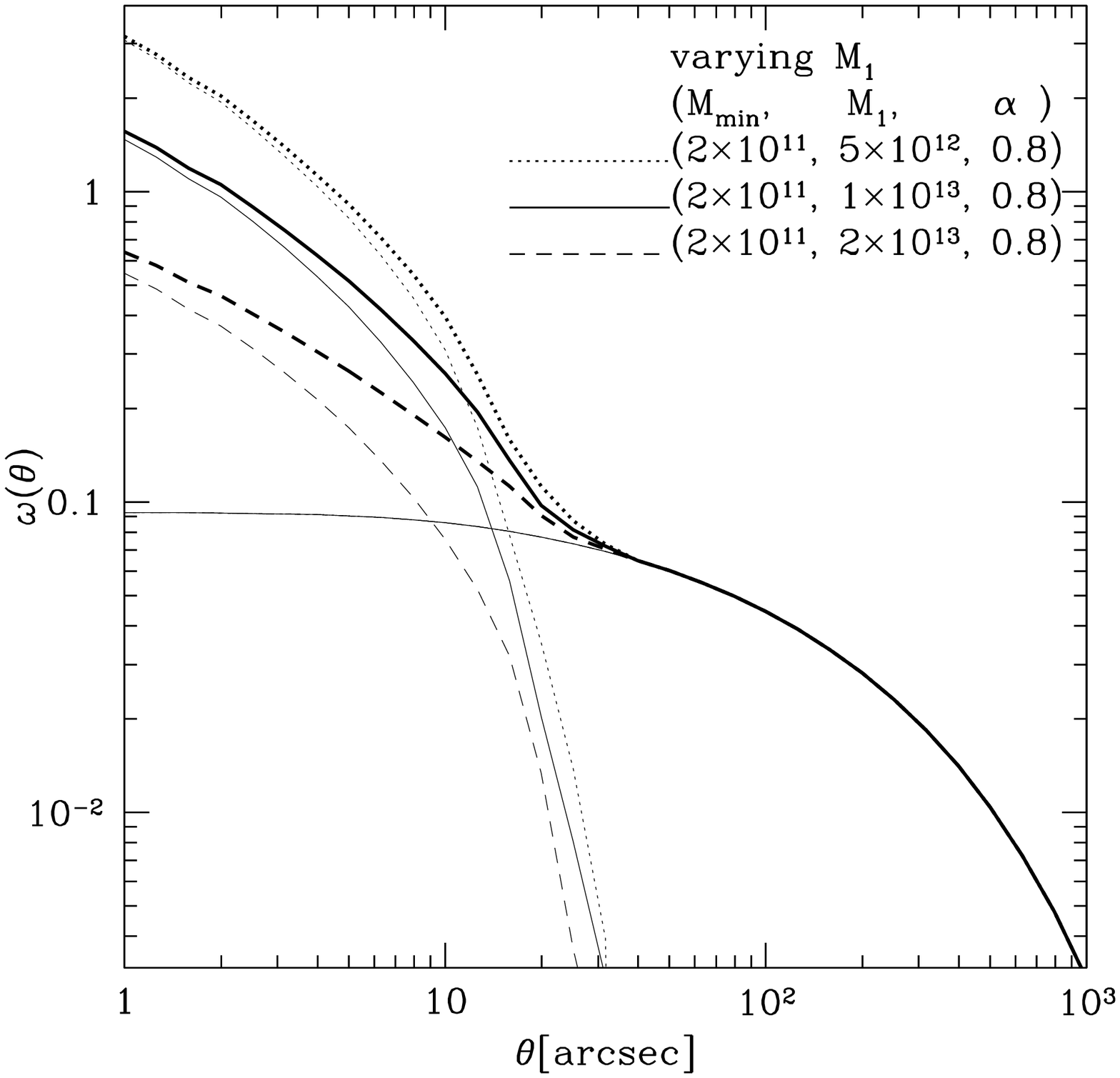}
\end{minipage}
\end{center}
\caption{Dependence of varying $M_1$ on the shape of angular two-point 
correlation functions.
HOF parameters for three cases are inserted in the plot. 
Two thin lines show 1-halo (dominates on small scales) and 
2-halo term (dominates on large scales) and thin lines show the sum of them.
We take the selection function of LBGz4 to compute those correlation functions 
for an illustrative purpose.}  
\label{fig:varying-M1}
\end{figure}

\begin{figure}
\begin{center}
\begin{minipage}{8.4cm}
\epsfxsize=8.4cm 
\epsffile{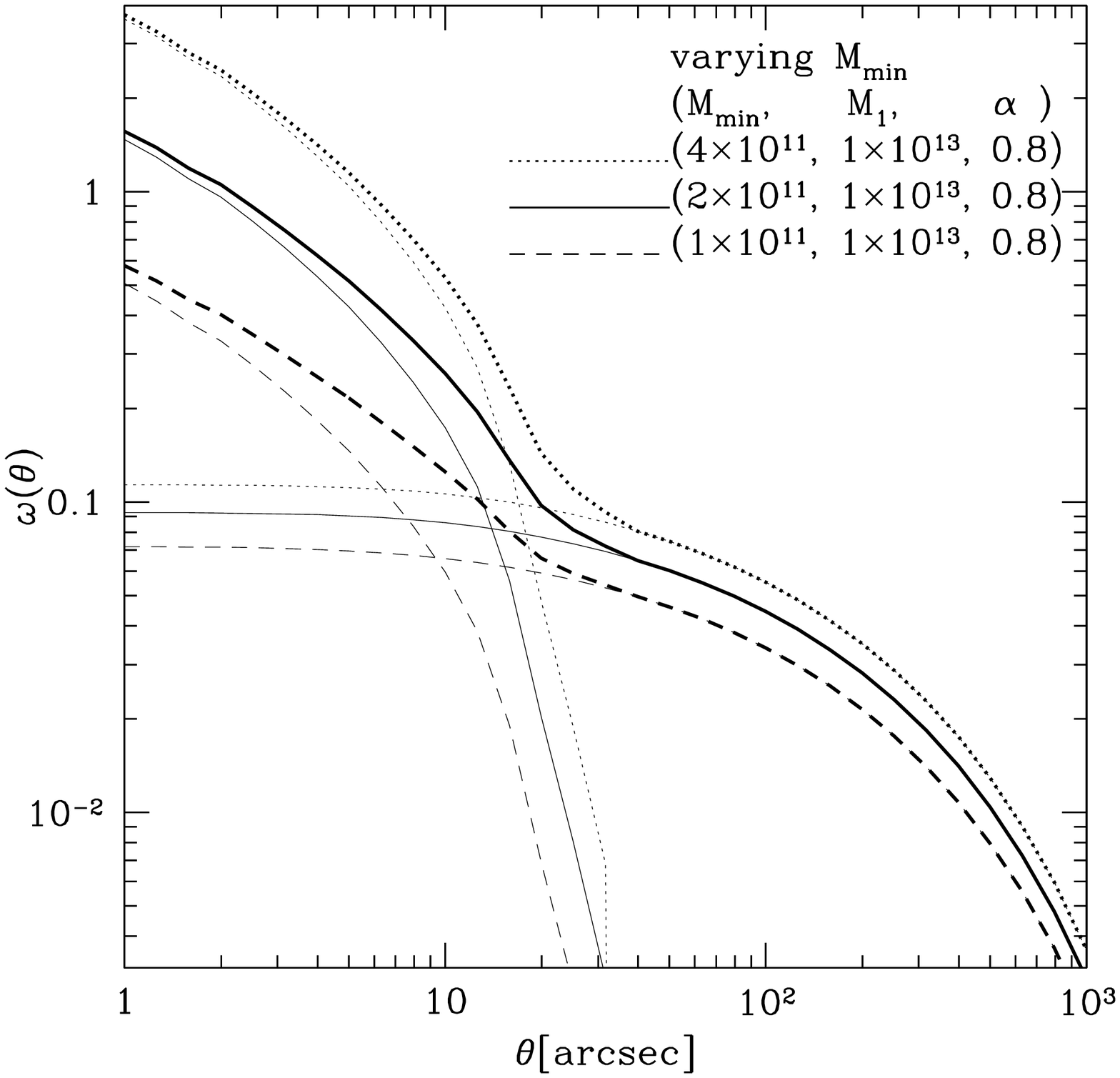}
\end{minipage}
\end{center}
\caption{Same as Figure \ref{fig:varying-M1}, but for $M_{min}$.}  
\label{fig:varying-Mmin}
\end{figure}

\begin{figure}
\begin{center}
\begin{minipage}{8.4cm}
\epsfxsize=8.4cm 
\epsffile{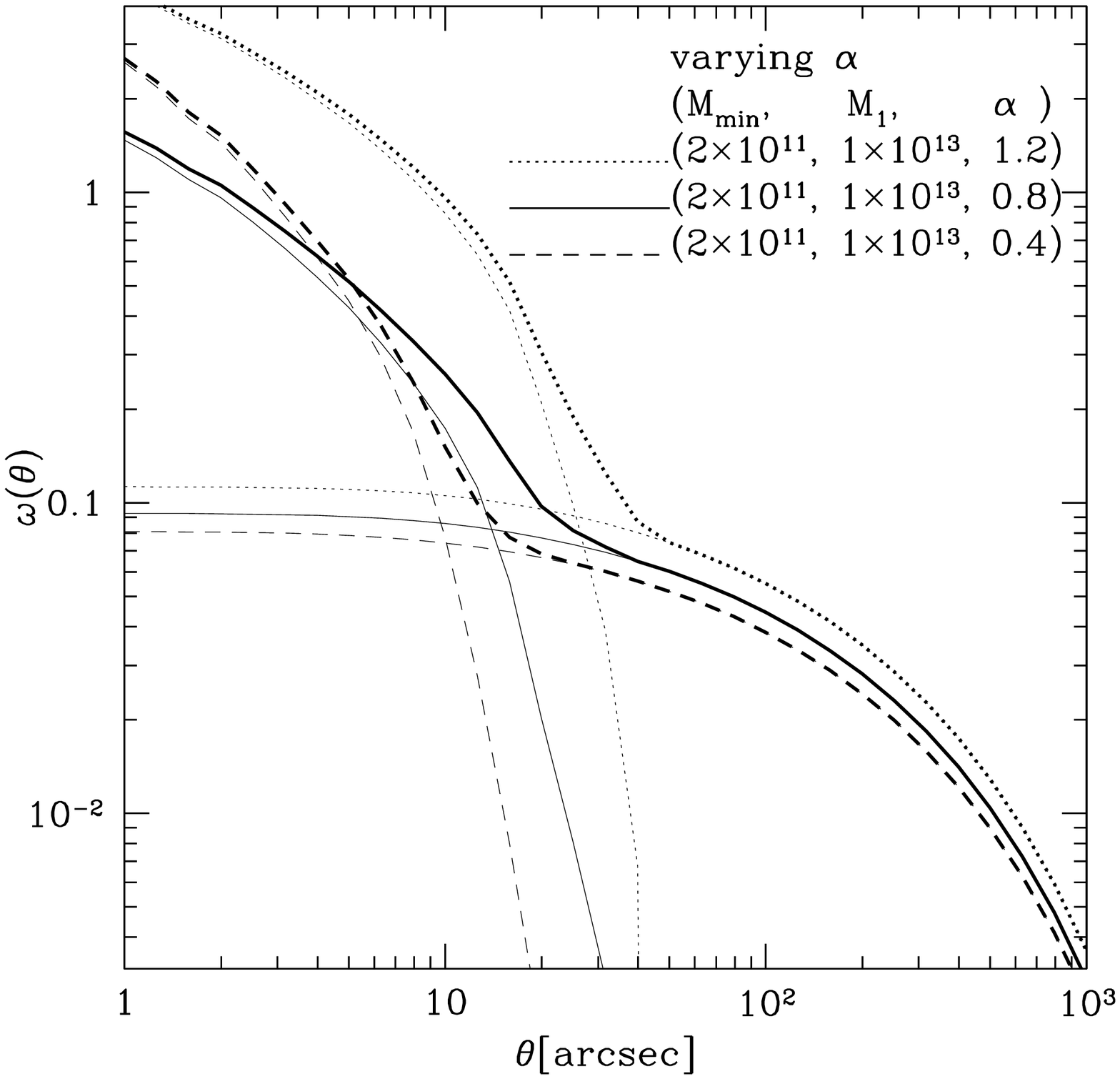}
\end{minipage}
\end{center}
\caption{Same as Figure \ref{fig:varying-M1}, but for $\alpha$.} 
\label{fig:varying-alp}
\end{figure}

Here we summarize major characteristics of changes in the shape of 
the two-point correlation functions made by varying HOF parameters.
Figure \ref{fig:varying-M1}-\ref{fig:varying-alp} show the angular 
correlation functions for variety of HOF models.
Each Figure demonstrates the effect of varying one HOF parameter with 
fixing other two parameters. 
In all plots, the selection function of LBGz4 is adopted for an illustrative
purpose.
It should be noticed that a degree of change in the shape of the correlation 
function made by varying one parameter depends on a choice of other 
two parameters, thus the plots should be understood as illustrative 
examples, and we just focus on major characteristics from a qualitative 
point of view.

Figure \ref{fig:varying-M1} shows the effect of varying $M_1$ on $\omega$.
Since the 2-halo term does not depend on $M_1$, the correlation function
on larger scales is not affected by the change of $M_1$.
The amplitude of the 1-halo term decreases with increasing $M_1$.
This is mainly due to the decrease in the contribution from smaller mass halos.

Figure \ref{fig:varying-Mmin} is for effect of varying $M_{min}$.
The amplitude of the 2-halo term increases with $M_{min}$, 
because of a larger bias factor for more massive halos.
The amplitude of the 1-halo term also increases with $M_{min}$.
This is largely the decrease in the contribution from smaller mass halos.

Finally Figure \ref{fig:varying-M1} shows the effect of varying $\alpha$ 
on $\omega$.
Varying $\alpha$ changes the fraction of galaxies in massive halo to less  
massive halos. A larger $\alpha$ gives more weight for galaxies in 
massive halos.
Since a stronger bias factor for more massive halo, a larger $\alpha$ 
gives a larger amplitude of the 2-halo term.
The change in the slope of the 1-halo term with varying $\alpha$ 
is explained as follows:
Since, roughly speaking, galaxies in larger halos contribute to 
the 1-halo term on relatively larger scales, while galaxies in smaller 
halos can only contribute to the 1-halo term on smaller scales.
Varying $\alpha$ changes the fraction in the contribution from larger halos
to smaller mass halos, and accordingly changes the slope of the 1-halo term.
A smaller $\alpha$ results a steeper slope, because of a more weight for 
galaxies in smaller halos.

\label{lastpage}

\end{document}